\documentclass[sigconf,nonacm, authorversion]{aamas}

\usepackage{balance} 

\usepackage[ruled,vlined]{algorithm2e}
\usepackage{amsmath}
\usepackage{booktabs}
\usepackage{color}
\usepackage{caption}
\usepackage{bmpsize}
\usepackage{tcolorbox}

\newcommand{\ourage}{{\sf AIM-RM}\xspace}
\newcommand{\llm}{\mathrm{LLM}\xspace}
\newcommand{\invagent}{{\sf InvAgent}\xspace}
\newcommand{\opt}{{\sf Opt}\xspace}
\newcommand{\ippo}{{\sf IPPO}\xspace}
\newcommand{\mappo}{{\sf MAPPO}\xspace}
\newcommand{\basestock}{{\sf Base-Stock}\xspace}
\newcommand{\tracking}{{\sf{Tracking-Demand}}\xspace}
\newcommand{\stepp}{$\mathcal{P}_{\text{SD}}$\xspace}
\newcommand{\memoryp}{$\mathcal{P}_\text{MU}$ \xspace}
\newcommand{\safestockp}{$\mathcal{P}_{\text{SS}}$\xspace}
\newcommand{\decisionprompt}{
\mathcal{P}_{\text{DM}}
}
\newcommand{\hookdownright}{\hookrightarrow}
\newtheorem{remark}{Remark}

\setcopyright{ifaamas}
\acmConference[AAMAS '26]{Proc.\@ of the 25th International Conference
on Autonomous Agents and Multiagent Systems (AAMAS 2026)}{May 25 -- 29, 2026}
{Paphos, Cyprus}{C.~Amato, L.~Dennis, V.~Mascardi, J.~Thangarajah (eds.)}
\copyrightyear{2026}
\acmYear{2026}
\acmDOI{}
\acmPrice{}
\acmISBN{}

\title[AI Agent Systems for Supply Chains: Structured Decision Prompts and Memory Retrieval]{AI Agent Systems for Supply Chains: Structured Decision Prompts and Memory Retrieval}

\thanks{This paper is a full version of the extended abstract accepted by the 25th International Conference
on Autonomous Agents and Multiagent Systems (AAMAS 2026)}

\author{Konosuke Yoshizato}
\affiliation{
  \institution{National Institute of Advanced Industrial Science and Technology}
  \city{Tsukuba}
  \country{Japan}
  }
\affiliation{
  \institution{Nagoya Institute of Technology}
  \city{Nagoya}
  \country{Japan}
  }
\email{yoshizato.konosuke@otsukalab.nitech.ac.jp}

\author{Kazuma Shimizu}
\affiliation{
  \institution{National Institute of Advanced Industrial Science and Technology}
  \city{Tsukuba}
  \country{Japan}
}
\affiliation{
  \institution{NEC Corporation}
  \city{Tokyo}
  \country{Japan}
  }
  \email{smzkzm2019@nec.com}

\author{Ryota Higa}
\affiliation{
  \institution{National Institute of Advanced Industrial Science and Technology}
  \city{Tsukuba}
  \country{Japan}
}
\affiliation{
  \institution{NEC Corporation}
  \city{Tokyo}
  \country{Japan}
  }
\email{r-higaryouta@nec.com}

\author{Takanobu Otsuka}
\affiliation{
  \institution{National Institute of Advanced Industrial Science and Technology}
  \city{Tsukuba}
  \country{Japan}
}
\affiliation{
  \institution{Nagoya Institute of Technology}
  \city{Nagoya}
  \country{Japan}
  }
\email{otsuka.takanobu@nitech.ac.jp}

\begin{abstract}
This study  investigates large language model (LLM) -based multi-agent systems (MASs) as a promising approach to inventory management, which is a key component of supply chain management. Although these systems have gained considerable attention for their potential to address the challenges associated with typical inventory management methods, key uncertainties regarding their effectiveness persist. Specifically, it is unclear whether LLM-based MASs can consistently derive optimal ordering policies and adapt to diverse  supply chain scenarios. To address these questions, we examine an LLM-based MAS with a fixed-ordering strategy prompt that encodes the stepwise processes of the problem setting and a safe-stock strategy commonly used in inventory management. Our empirical results  demonstrate that, even  without detailed prompt adjustments, an LLM-based MAS can determine optimal ordering decisions in a restricted scenario. To enhance adaptability, we propose a novel agent called \ourage, which leverages similar historical experiences through similarity matching. Our results  show that \ourage outperforms benchmark methods across various supply chain scenarios, highlighting its robustness and adaptability.
\end{abstract}


\newcommand{\BibTeX}{\rm B\kern-.05em{\sc i\kern-.025em b}\kern-.08em\TeX}

\begin{document}

\newpage
\maketitle

\section{Introduction}

Multi agent systems (MASs)  are intelligent systems that autonomously perceive their environment, make decisions, and execute actions. 
These systems have gained significant attention in many actual applications.
However, for such a system to operate effectively, multiple AI agents must act independently and collaboratively. 
In this context, recent progress in large language models (LLMs) has greatly enhanced the abilities of AI agents, making them highly suitable for diverse industrial applications, including manufacturing, finance, and customer service ~\cite{quan2024invagent,zhang2024multimodal,li-etal-2024-cryptotrade,liang2025llm}. 
In particular, LLM-based MASs can address the complex interactions and negotiations involved in real-world applications.

Supply chain management (SCM) is a particularly promising domain for applying LLM-based MASs. 
Modern supply chains face mounting complexities, such as globalization \cite{e4sssupply, supplychainannualrepo} and climate change \cite{park2025predictive}.
At the core of these challenges is inventory management, which is essential for procurement, production, and distribution throughout the supply chain.
Applying an MAS to inventory management enables individual companies to make autonomous decisions while facilitating efficient coordination among agents across the supply chain\cite{farazi2025enhancing, kaihara2006multi, koketsu2022concept, moyaux2003multi}. This approach supports optimization at the individual firm level and fosters collective benefits for all stakeholders, which is crucial for addressing SCM issues \cite{jia2023novel}. 
\begin{figure}[t!]
  \includegraphics[width=0.46\textwidth]{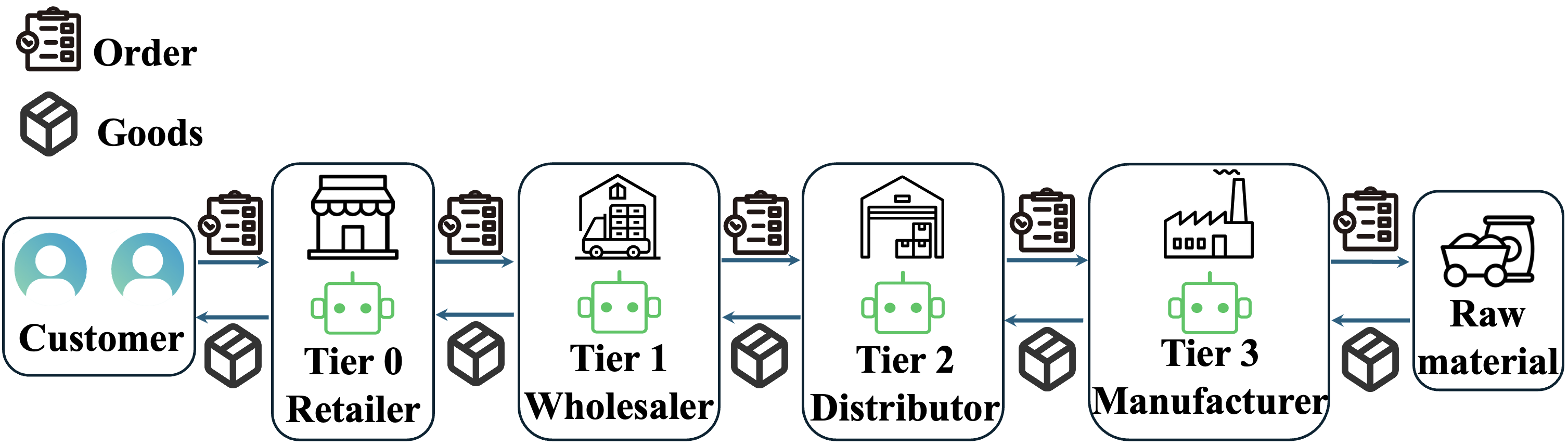}
  \caption{ Multi agent system for inventory management in a supply chain. At each tier, the agent places orders with the upstream agent and receives goods after a delivery lead time.}
  \label{fig:sc}
  \Description{supplychain}
\end{figure}

Existing research on inventory management predominantly employs one of two main approaches. 
The first approach uses heuristics, such as the base-stock policy. Although these heuristics are widely used, they often require problem-specific, laborious customization to achieve high performance \cite{geevers2024multi}.
The second approach employs reinforcement learning (RL), enabling systems to autonomously develop flexible and effective strategies. 
However, RL, particularly in multi-agent and complex supply chain contexts, 
incurs significant computational costs, making it impractical for real-world applications \cite{boute2022deep, ziegner2025iterative}. 
Therefore, applying this approach to actual supply chain operations remains challenging.

Recent research has increasingly focused on applying LLM-based AI agents to SCM \cite{li2023large,quan2024econlogicqa,jannelli2024agentic, kirshner2024talking,wang2025llms}. 
Many of these studies address inventory management problems by leveraging knowledge sharing and domain knowledge in manufacturing through retrieval-augmented generation (RAG) \cite{lewis2020retrieval}. 
However, recent advancements in LLMs, such as GPT-5 \cite{openai2025gpt} and o4-mini \cite{openai2025o3o4}, have demonstrated strong reasoning and numerical abilities. 
These capabilities suggest that state-of-the-art models can perform efficient inventory management in supply chains by adopting effective decision-making policies without relying on external information sources. 
Therefore, the ability of current LLM foundation models to independently  handle inventory management tasks must be evaluated, 
as this evaluation represents a critical first step in exploring the  potential of AI agents for supply chain optimization.

Our first goal is to investigate the current limitations of LLM-based MASs. 
Understanding these constraints is crucial for our second objective: designing adaptive LLM-based MASs that can operate in diverse supply chain situations.
A key challenge is that such systems often rely on situation-specific prompts, which can cause unstable outputs and unreliable performance \cite{zhoubatch,salinas2024butterfly, sclarquantifying, ngweta2025towards}. 
Furthermore, the process of prompt-tuning process often requires considerable effort \cite{liu2023pre,dong-etal-2025-model}, making it potentially impractical. 
To address these obstacles, we propose a method that integrates a procurement strategy and historical transaction data. 
This method is designed to improve context awareness and collaborative decision-making, 
enabling LLM-based MASs to overcome the challenges faced by traditional approaches and improve optimization in real-world SCM.

\subsection{Our Contributions }

This study had two main goals. First, we investigated the current limitations of LLM-based MASs in the field of inventory management. 
To achieve this goal, we construct experimental scenarios that simulate various supply chain situations and examine the performance of MASs . 
Our experiments reveal that under specific conditions, 
an LLM-based MAS in the supply chain successfully identifies and implements an optimal ordering policy without necessitating prompt adjustments tailored to the scenario. 
To the best of our knowledge, this is the first study to demonstrate that an LLM-based MAS can achieve optimal solutions for multi-echelon inventory management problems. 
However, the results also show that an LLM-based MAS that reaches the optimal ordering policy fails to identify optimal policies in different supply chain scenarios, 
indicating a limitation of LLM-based MASs. 

Second, we aim to design an adaptive LLM-based MAS that performs robustly across diverse supply chain situations. To achieve this goal, we propose a method that enables agents to retrieve and reference past experiences similar to the current context. 
In particular, by storing RL results as past experiences, our proposed MAS implicitly coordinates distributed agents and provides coherent, system-wide behavior, 
enabling an LLM-based MAS to adapt flexibly to diverse supply chain scenarios. 
Consequently, our study identifies both the capabilities and limitations of an LLM-based MAS in inventory management 
and introduces an approach to improve adaptability.

\section{Literature Review}
The inventory management problem in multi-layer supply chains is modeled as a multi-echelon inventory management problem 
and has been extensively studied over many years (see \cite{de2018typology} for details). 
Some of the existing research have modeled this problem as a type of Markov decision process (MDP) and derived optimal policies using dynamic programming \cite{sarimveis2008dynamic}. 
However, owing to  the complexity of the problem and the challenges in handling mathematical models, heuristic methods such as base-stock policy are commonly used in practice \cite{geevers2024multi, ziegner2025iterative}.

The multi-echelon inventory management problem has been implemented in the OpenAI OR-gym simulation library \cite{HubbsOR-Gym}, facilitating the application of multi-agent RL  methods. 
Numerous  studies in this area have focuses on the centralized training with decentralized execution (CTDE) architecture \cite{kaihara2006multi}, 
which is intermediate between fully centralized and fully decentralized models, allowing agents to access global information during training while executing learning policies in a decentralized manner.  
Although CTDE and related RL methods \cite{de2020independent,liu2022multi,mousa2024analysis,geevers2024multi} often exhibit strong performance, their practical application remains limited by challenges such as high sample complexity and the necessity to construct simulators calibrated with real data \cite{madeka2022deep}.

In recent years, research on the application of LLM-based MASs to multi-echelon inventory management problems has gained significant momentum. 
The study  \cite{quan2024invagent} proposed \invagent and demonstrated that it can adapt its decisions to various demand scenarios 
compared to the heuristic and RL methods. 
The results showed that a simple LLM-based MAS enables efficient SCM.
Subsequent studies have examined effective decision-making with human-like negotiation and communication among agents \cite{kirshner2024talking, jannelli2024agentic}. 
Beyond leveraging only foundation LLMs, the study  \cite{wang2025llms} proposed augmenting agent capabilities with external knowledge sources through RAG , 
using relevant materials, such as textbooks and literature on manufacturing. 
Although these studies have significantly advanced the integration of LLM-based agents with inventory management in supply chains, 
it remains unclear whether simple LLM-based agents can achieve truly optimal policies.  
Furthermore, the fundamental limitations of these agent systems in complex inventory management situations have yet to be fully elucidated.

A recent study has indicated that increased reasoning effort in LLMs does not always result in enhanced task performance \cite{ghosal2025does}. This counterintuitive phenomenon, called {\it overthinking}, occurs when the number of reasoning tokens increases beyond a certain threshold.
This property of reasoning models is critical, as it appears to limit the effectiveness of LLM-based MASs in executing complex tasks such as multi-echelon inventory management problems.

\section{Methodology}
In this section,  
we first describe the supply chain inventory management problem considered in this study and then present our model.

\subsection{Multi-Echelon Inventory Management}
\label{sec:echelon}

\begin{figure}[h]
  \includegraphics[width=0.45\textwidth]{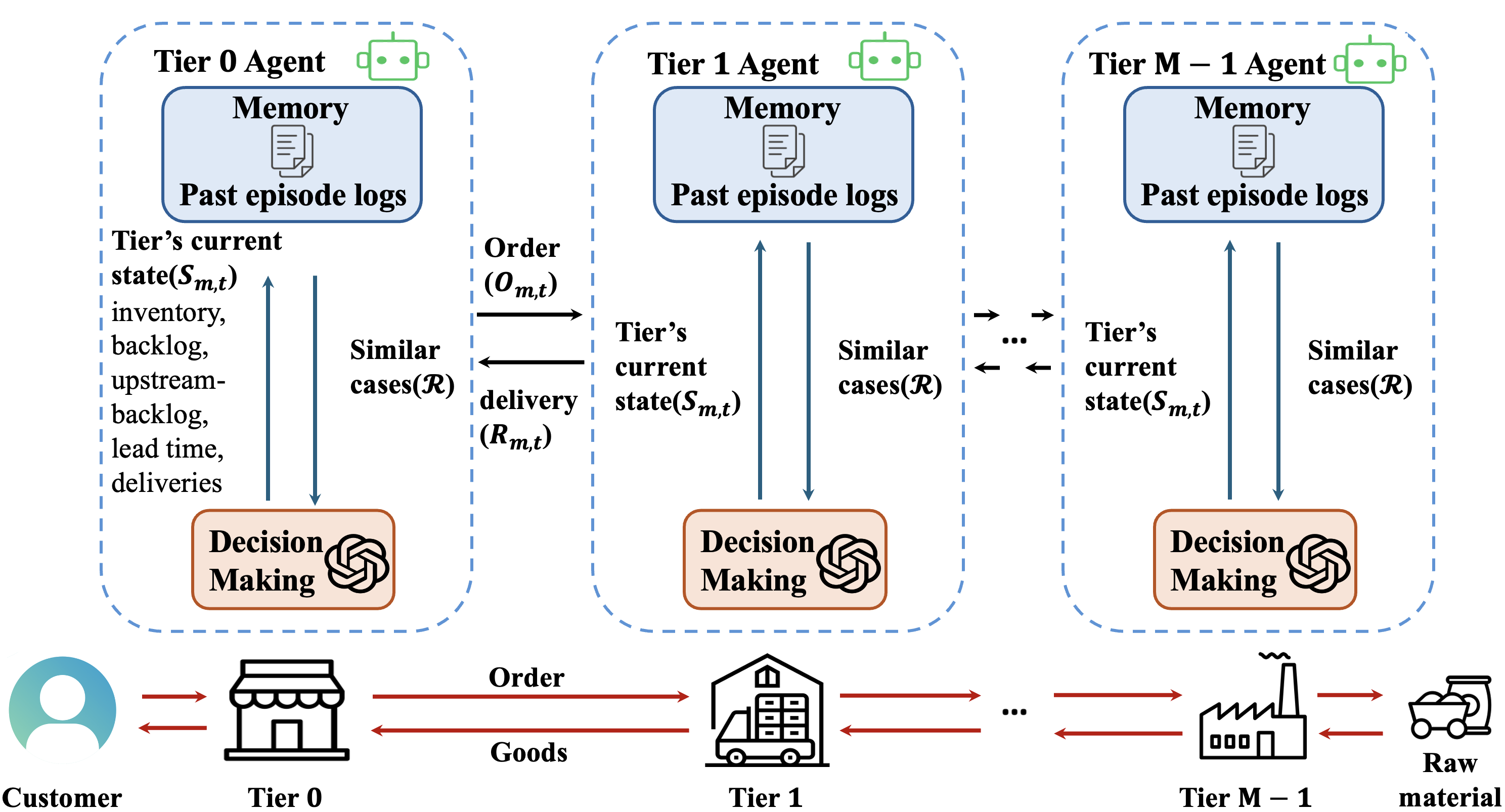}
  \caption{Framework of the interaction between agents and the supply chain environment. Starting with the downstream agent, the agent at each tier observes its state and places an order with the upstream agent in a round. Our proposed agent retrieves historical experience data through similarity matching and then uses them when placing an order.}
  \label{fig:env}
  \Description{Environment}
\end{figure}

The problem of inventory management in multi-tier supply chains is modeled as a multi-echelon inventory management problem.
In this model, each tier in the supply chain stores materials and uses them for production. 
The tiers are labeled in ascending order, with tier $0$ representing the retailer. 
At each tier, producing one unit of the product requires one unit of material, with the maximum production quantity per period limited by facility capacity $c_m$. 
Finished products cannot be stored as inventory and are immediately shipped to the downstream tiers. 
The products manufactured at tier $m$ are shipped to tier $m+1$ and arrive after a predetermined transportation lead time $L_m$. 
Upon arrival, these products become inventory at tier $m+1$ and are used for subsequent production. 
Figure \ref{fig:env} illustrates the supply chain flow.

Next, we explain the procedure of operation within an episode.
The episode comprises $T$ periods. In each period $t \in [T] = \{1, \dots, T\}$, 
the agent at each tier sequentially executes the following four steps:

(i) Inventory Replenishment: The materials ordered $L_m$ periods earlier are delivered, and the inventory level is replenished.

(ii) Ordering and Demand Observation: At each tier $m \in M$, orders for $O_{m,t}$ units of materials are placed with suppliers. 
Simultaneously, the demand from buyers is observed and fulfilled with the available inventory as much as possible.

(iii) Shipping Products: Each agent determines its shipment quantity $R_{m,t}$, which is limited by either the production capacity or available inventory, whichever is lower. After shipment, the remaining inventory is $I_{m,t}$. 
Any unfulfilled demand $B_{m,t}$ results in a backlog, which incurs opportunity costs and is prioritized for fulfillment in future periods.

(iv) Profit and Loss Calculation: Each agent calculates its profit and loss $P_{m,t}$ by aggregating the product sales revenue, material procurement costs, backlog penalties, and inventory holding costs.

The above quantities can be formulated as follows \cite{HubbsOR-Gym}: 
\begin{align}
\label{eq:replenish}
R_{m,t} &= \min(
B_{m+1, t-1} + O_{m,t}, c_{m+1}, I_{m+1, t-1} 
\\
&\quad\quad + R_{m+1, t-L_{m+1}}), m\in\mathcal{M},
\\
\label{eq:retreplenish}
R_{M-1,t} &= O_{M-1,t}, 
\\
\label{eq:sales}
S_{m,t} &= R_{m-1,t}, 
\\
\label{eq:retsales}
S_{0,t} &= \min\left(B_{0,t-1} + D_{t}, c_{0}, I_{0,t-1} + R_{0,t-1}\right)
\\
\label{eq:backlog}
B_{m,t} &= B_{m,t-1} + O_{m-1,t} - S_{m,t}
\\
\label{eq:retbacklog}
B_{0,t} &= B_{0,t-1} + D_t - S_{0,t}
\\
\label{eq:inv}
I_{m,t} &= I_{m,t-1} + R_{m,t-L_m} -S_{m,t}
\\
\label{eq:profit}
P_{m,t} &= P_mS_m - r_mR_{m,t} - k_mB_{m,t} -h_m I_{m,t} 
\end{align}

\subsection{Agents' Behavior in Multi Echelon SCM}
\label{sec:agentbehavior}
In this problem setting, the agent at each tier selects an action $a_{m,t}=O_{m,t}$, which influences the environment and leads to changes in variables, such as rewards and inventory levels. This process can be modeled as an MDP . The interaction between the agents and the environment is implemented using a user proxy. The procedure is as follows:

(i) The environment is initialized at the beginning of the first round. 

(ii) The agent at each tier observes the state of the current round from the environment.

(iii) Upon observing its state $s_{m,t}$, each agent decides its action $a_{m,t}$ and sends it to the user proxy.

(iv) The user proxy transmits each agent's action to the environment, which then returns the next state $s_{m,t+1}$ and the reward for the round.

(v) The user proxy delivers the updated state and reward to each agent. When the current round has not reached the final round $T$, the process returns to step (ii).
The state $s_{m,t}\in \mathbb{R}^{4 + 2L_{m}}$ is defined as  
\begin{align}
s_{m,t} 
&=[I_{m,t-1}, B_{m, t-1}, B_{m+1, t-1}, S_{m,t-L_{m}}, \dots, S_{m,t-1}, R_{m,t-L_{m}},\dots,
\nonumber \\
& R_{m,t-1}].  
\end{align}

\subsection{Proposed Agent}
Here, we propose \ourage, an AI agent that can leverage insights from past similar situations. This agent comprises two main modules: 
memory  and decision-making modules. The details of each module are described below. 

This module  determines ordering actions using the decision prompt $\decisionprompt$, which is the base prompt for decision-making and contains the necessary information, including the current round $t$, tier number $m$, current state $s_{m,t}$, historical data $\mathcal{H}$, and demand description $\mathcal{D}$. 
The historical data $\mathcal{H}$ are used only when the memory module introduced in this section is available.   
The demand description $\mathcal{D}$ describes the customer demand across all rounds.
For illustration, Appendix~\ref{appendix:prompts} provides an example of a decision prompt based on the prompt proposed by \cite{quan2024invagent}. 
Upon receiving this prompt, the decision-making module generates not only an action but also the reasoning behind each decision, as formalized by
\begin{align}
DE_{m,t} = \llm_{\text{DM}}(
\decisionprompt (s_{m,t},\mathcal{H}, \mathcal{D})).
\end{align}
Here, the decision output $DE_{m,t}$ comprises the order quantity $O_{m,t}$ and the reason.

To facilitate the derivation of optimal policies by an LLM, we tune the decision prompt $\decisionprompt$. However, the initial version proposed in \cite{quan2024invagent} does not explicitly include critical information about the multi-echelon inventory management process, such as order arrivals and order timing. This omission may reduce the accuracy of the order quantity decisions made by the LLM.
To address this limitation, we revised $\decisionprompt$ by incorporating a step-by-step description \stepp~ that clearly outlines the procedures defined in Section~\ref{sec:echelon}. This modification was intended to improve the LLM's understanding of the inventory management process and support more accurate decision-making. To integrate the step-by-step description into the decision prompt, \stepp iswas inserted immediately after $\decisionprompt$. As a reference, an example of the step-by-step description \stepp~ is provided in Appendix~\ref{appendix:prompts}

Although the step-by-step description \stepp~ provides an overview of the environment, it does not specify concrete ordering strategies. To address this, we add a safety stock strategy to $\decisionprompt$, which determines the order quantity by considering both the lead time and forecasted demand and incorporates safety stock (SS). 
First, we define the scheduled inventory increase as $\tilde{I}_{m,t} = I_{m,t-1} +\sum_{\tau=t-L_{m}}^{t-1}R_{m,\tau} - B_{m,t-1}$, which represents the increase in the inventory level over the next lead time. Then, we define the safety stock and future consumption as $SS = z \hat{\sigma}\sqrt{L_m + 1}$ and $\tilde{C}_{t,m} = (L_m + 1)\hat\mu + SS$, respectively. Here, $\hat\mu$ is the forecasted demand, $\hat{\sigma}$ is the demand variance, and $z$ is the safety factor, determined by the target service level. This safety stock formula is a traditional formula used in inventory management \cite{gonccalves2020operations}. 
The order quantity is then determined based on the gap between $\tilde{D}_{t,m}$ and $\tilde{C}_{t,m}$. 
To incorporate the safety stock strategy into the decision-making process, \safestockp is inserted after \stepp.
The prompt example for this safety stock strategy is provided in Appendix~\ref{appendix:prompts}.

To ensure robust performance across diverse scenarios without the need for extensive prompt tuning, we introduce an external implementation of memory $\mathcal{M}$ that stores history, where $\mathcal{M}$ is a set of $(s_{m,t}, O_m, P_m)$. This memory is maintained in a vector database, which enables the retrieval of similar past situations based on state vector similarity. To further enhance the effective use of historical experiences, we incorporate a prompt specifying how to utilize this memory into the decision prompt $\decisionprompt$ by inserting a memory usage prompt \memoryp after \stepp.
For illustration, an example of \memoryp is presented in Appendix~\ref{appendix:prompts}. 

\ourage can store not only historical episodes but also simulation results for the multi-echelon inventory management problem in its history, enabling indirect coordination within a distributed architecture and yielding coherent, system-wide behavior. 
For example, in an RL simulation where each agent's optimal action is specified, storing each agent's portion of the result enables them to select actions that are implicitly coordinated with the other agents' actions through the stored result.

To efficiently leverage the memory function, a similarity search using vector representations is performed by calculating the Euclidean distances between the current state vector and the state vectors of historical experiences. If the computed distance is below a predefined threshold, the corresponding past episode is considered to be similar and selected for reference. Furthermore, up to $K$ of the closest matches can be retrieved and utilized by the agent.

\begin{remark}
{\rm 
The computational cost of the similarity matching process is estimated to be $\mathcal{O}(|\mathcal{M}|d + X + K)$, where $|\mathcal{M}|$ is the cardinality of memory $\mathcal{M}$, $d$ is the dimension of the embedding space, $X$ is the computational cost of sorting, and $K$ is the neighbor count. The $\mathcal{O}(|\mathcal{M}|d)$ term results from computing the Euclidean distance between each vector in $\mathcal{M}$ and the current state. The $\mathcal{O}(K)$ term arises from identifying memories with similarities that are below the threshold. The computational cost $X$ can be $\mathcal{O}(|\mathcal{M}|\log(|\mathcal{M}|))$ when we use some quicksort algorithms, whereas it can be $\mathcal{O}(|\mathcal{M}|^2)$ in the worst case. Overall, the similarity matching process requires a low computational cost as long as $M$ and $d$ are not simultaneously large. 
}
\end{remark}

\begin{algorithm}[t!]
\caption{One-round decision procedure for all stages in a period}
\label{alg:one_round_per_period}
\DontPrintSemicolon
\SetAlgoLined
\SetKwInOut{Input}{Input}
\SetKwInOut{Output}{Output}
\SetKwInOut{Data}{Data}

\Input{current period $t$, environment $\mathrm{Env}$, current observable state $s_{m,t}$, scenario $\texttt{scenario}$}
\Output{dictionary of actions $\texttt{action\_dict}$}
\Data{embedding map $\phi:\mathcal{S}\to\mathbb{R}^d$; per-stage memory $\{\mathcal{M}_m\}$; threshold $\tau$; neighbor count $K$; decision prompt $\decisionprompt$; demand description $\mathcal{D}$
}

Set $\texttt{action\_dict} \gets \{\}$\;

\BlankLine
\For{$m \gets 0$ \KwTo $M - 1$ }{

  Select the $K$ nearest neighbors $\mathcal{N}_K$ from $\mathcal{M}_m$ according to the Euclidean distance $d = \lVert \phi(s_{m,t}) - v \rVert_2$ in ascending order, where $v$ is the state embedding vector of $\ell \in \mathcal{M}_m$ \;

  Filter similar cases $\mathcal{R}$ by applying the threshold $\tau$:  
    $\mathcal{R} = \{ n \in \mathcal{N}_K \mid d_i < \tau\}$.\;
  Obtain the output from the decision-making LLM  with $\decisionprompt$, $\mathcal{R}$, $\mathcal{D}$ and $s_{m,t}$:\\
    $(O_{m,t},\ \mathrm{reason}) \gets \llm_{\text{DM}}(
\decisionprompt (s_{m,t},\mathcal{R}, \mathcal{D})).$\;
  Submit the action $O_{m,t}$ to the environment and observe the next state and reward:
    $(s_{m,t+1},\ P_{m,t}) \gets \mathrm{Env}(m,\ O_{m,t})$.\;
  Record the action for this stage: $\texttt{action\_dict}[m] \gets O_{m,t}$.\;
  Update the memory with the current record:
    $\mathcal{M}_m \gets \mathcal{M}_m \cup \{(\boldsymbol{v}_{m,t},\ O_{m,t},\ P_{m,t})\}$.\;
}
\BlankLine
\Return $\texttt{action\_dict}$\;
\end{algorithm}

\section{Experiments}
 
In this section, we first describe the experimental setup for the multi-echelon inventory management problem considered in this study. Next, we introduce the baseline methods used for comparison. Finally, we present the results of the numerical experiments, including an analysis of the total rewards.

\begin{table}[t]
    \centering
        \caption{Parameters of the experimental scenarios. The values in the list correspond to the values of the stages 0 through 3, arranged sequentially from left to right.}
    \begin{tabular}{lcc}
        \toprule
        \textbf{Scenario} & \textbf{Uniform} & \textbf{Diverse} 
        \\
        \midrule
        Number of Stages
        & 4 & 4  
        \\
        Number of Periods
        & 12 & 12  
        \\
        Initial Inventories
        & [12,12,12,12]  & [12,14,16,18] 
        \\
        Lead Times
        & [2,2,2,2]  & [1,2,3,4]
        \\
        Product Capacities      & [20,20,20,20] &[20,22,24,26] 
        \\
        Sales Prices            & [0,0,0,0]     & [9,8,7,6]     
        \\
        Order Costs             & [0,0,0,0]      & [8,7,6,5]    
        \\
        Backlog Costs           & [1,1,1,1]     & [1,1,1,1]    
        \\
        Holding Costs           & [1,1,1,1]     & [1,1,1,1] 
        \\
        \bottomrule
    \end{tabular}
    \label{tab:scenario_parameters}
\end{table}

\subsection{Experimental Setting}

We assign the same agent to each tier in the multi-echelon inventory management problem to construct a multi-agent system.
To ensure that each agent is aware of its respective tier, we provide a system prompt as shown in \cite{quan2024invagent}, which is given in Appendix~\ref{appendix:prompts}.
We provide five experimental scenarios with different supply chain parameter settings and customer demand patterns. 
For the supply chain parameter configuration, we consider two settings: the {\it uniform} setting, where all tiers share identical parameters, such as the initial inventory and lead time, and the {\it diverse} setting, where these parameters differ across the tiers. 
The specific parameters used in the experimental scenarios are listed in Table \ref{tab:scenario_parameters}. 
For customer demand patterns, we use three types of patterns: {\it constant}, where the demand is fixed at $4$ in every round; {\it increasing}, 
where the demand grows with each round according to $D_{\text{inc}} = 2 + \lceil t/3 \rceil$; and {\it decreasing}, 
where the demand falls as $D_{\text{dec}} = 2 + \lceil (12-(t-1))/3 \rceil$. Here, $\lceil x \rceil$ denotes the smallest integer that is greater than or equal to $x$. 
All the agents are informed of customer demand through $\decisionprompt$, 
where the demand functions are specified and the volume of demand that appears in each round is explained. 
For reference, the demand description in the prompt is presented in Appendix~\ref{appendix:prompts}.

For the experimental configurations, we provide four distinct agent configurations and two LLM parameter settings. 
The first two configurations do not employ the memory function: an agent uses the decision prompt $\decisionprompt$ integrating the step-wise description \stepp, labeled as \invagent(w/ step desc), while another agent uses $\decisionprompt$ containing both \stepp and the safety-stock strategy \safestockp, denoted as \invagent(w/ step and safety-stock strategy). 
The other two configurations involve the memory function: one model does not preload any prior historical logs but stores historical logs during the episode, labeled as \ourage (w/o RL log),
and the other model is initialized with historical logs derived from the evaluation data from \ippo, denoted as \ourage(w/ RL log). 
\ourage(w/ RL log) retrieves the $K=6$ nearest neighbors to identify similar cases with the threshold value $\tau=2$. 
In both memory-based configurations, agents use the decision prompt $\decisionprompt$ with a step-wise description \stepp. In particular, for \ourage(w/ RL log), the prompt of each agent further contains a memory usage prompt \memoryp that indicates how to use the retrieved similar cases. 
For the LLM parameter configurations, all agents in each setting use OpenAI o4-mini, with the reasoning parameter set to either {\it high} or {\it medium}. 
To evaluate the performance of these configurations, we conduct $5$ episodes to evaluate the total rewards across total rounds and tiers; this number of episodes is determined to balance LLM usage costs, time consumption, and variability between episodes.

\subsubsection*{Comparison Targets}

\invagent\ is an agent that uses o4-mini as the decision-making module with two reasoning effort parameters {\it high} and {\it medium}. 
\invagent directly employs the prompt proposed in \cite{quan2024invagent}. 
Following the settings described in that reference, two variants are introduced: one incorporating the additional strategy (\invagent  w/ strategy) and the other without it (\invagent  w/o strategy). 
In addition, for the comparison between reasoning  and non-reasoning models, 
we provide two variants of \invagent that use GPT-4.1 instead of o4-mini.
Each condition was evaluated over $5$ independent trials.

For heuristic-based comparison targets, we include two representative heuristic methods: the base stock policy (\basestock) and the demand tracking policy (\tracking).
Under the base stock policy, the agent always orders enough units at each step to raise its inventory to a predefined upper limit, specifically ordering $(c_m - I_{m,t})$ units.
Under the demand tracking policy, the agent aims to maintain a target inventory level $\bar{I}_{t,m}$ based on the average sales over the maximum lead time period $\bar{S}$. At each step, it orders $\max \left\{\bar{I}_{m,t} - I_{t,m},0 \right\}$ units, where $\bar{S}_{m,t-1} = \sum_{u=1}^{L_{\text{max}}} S_{t-u}/L_{\text{max}}$ and $ \bar{I}_{t,m} = \bar{S}_{m,t-1} L_{m} + B_{m,t-1}.$ Here, $\bar{S}_{m,t-1}$ is the average sales over the maximum lead time period, and $\bar{I}_{t,m}$ is the corresponding target inventory.

As learning-based baselines under the CTDE framework, we use independent proximal policy optimization \ippo~\cite{de2020independent} and multi-agent proximal policy optimization \mappo~\cite{yu2022surprising}. 
\ippo~ independently trains each agent using the proximal policy optimization (PPO) algorithm \cite{schulman2017proximal}. This method employs parameter sharing across all agents, using a single shared policy to improve training efficiency.  
\mappo~ is a multi-agent variant of the PPO. This method employs a centralized value function conditioned on global observations from all agents, thereby reducing the non-stationarity that arises from concurrent multi-agent learning. 
To determine the hyperparameters for these models in each experimental scenario, we chose random combinations of hyperparameters from a candidate set, including hidden layer size, learning rate, and batch size, and we selected the combinations that achieved the highest score. Subsequently, we evaluated the average reward across 100 trials. The details of the hyperparameter selection are given in Appendix~\ref{appendix:reinforce}.

We can determine the optimal total reward across all tiers by solving the revenue maximization problem under the conditions \eqref{eq:replenish} - \eqref{eq:inv}. The optimal policy reward is denoted as \opt. We evaluate the performance in this experiment using the relative optimality gap $\Delta = |(\opt - r)/ \opt|$, where $r$ is the total reward of a policy. We provide the optimal values for all experimental scenarios in Appendix~\ref{appendix:numerical}.

\begin{remark}
{\rm 
A key difference between our experiment and that of \cite{quan2024invagent} is that we use a deterministic customer demand setting, eliminating randomness. This is because of the high cost of LLM usage, which makes it difficult to secure a statistically sufficient number of trials for scenarios involving random demand. We believe that our experimental setup remains adequate for evaluating whether the agents can successfully respond to demand trends. 
}
\end{remark}

\begin{table*}[t]
  \centering
  \caption{
  This table presents the average total reward for each model across episodes. The bolded numbers represent the highest scores among the models, excluding the reinforcement learning models (\ippo, \mappo). The rewards are expressed as the relative gap $\Delta = |(\opt - r)/ \opt|$, where $\opt$ is the optimal value and $r$ is the total reward. 
  The term {\it const-uni} in the first line stands for the constant-uniform setting. Similarly, the other names in the first line are abbreviations for different setting names. The first four models are the ones we have proposed.
  } 
  \label{tab:results_med}
  \centering
  \begin{tabular}{lccccc|c}
  \toprule
  \textbf{Model (Reasoning effort: Medium)} & \textbf{Const-Uni} & \textbf{Dec-Div} & \textbf{Dec-Uni} & \textbf{Inc-Div} & \textbf{Inc-Uni} 
  & \textbf{Average} 
  \\
  \midrule
  \invagent(w/ step desc) & 3.33 & 79.22 & 437.78 & 104.55 & 137.88 & 152.55
  \\
  \invagent(w/ step desc and SS strategy) & \textbf{0.00} & 115.06 & 315.56 & 264.88 & 425.00 & 224.10
  \\
  \ourage(w/o RL log) & 69.17 & 66.27 & 219.11 & 162.64 & 175.77 & 138.59
  \\
  \ourage(w/ RL log) & 60.00 & \textbf{56.02} & \textbf{171.11} & \textbf{95.12} & \textbf{74.09} & \textbf{91.27}
  \\
    \invagent (GPT-4.1, w/o strategy) & 120.00 & 244.58 & 991.11 & 474.79 & 262.88 & 418.67 \\
  \invagent (GPT-4.1, w/ strategy) & 76.67 & 88.25 & 428.89 & 133.06 & 181.81 & 181.74 \\
  \invagent (w/o strategy) & 46.67 & 100.30 & 202.22 & 171.90 & 93.18 & 122.85
  \\
  \invagent (w/ strategy) & 183.33 & 118.67 & 200.00 & 179.33 & 155.30 & 167.33
  \\
    \basestock & 146.67 & 140.36 & 340.00 & 162.81 & 112.12 & 180.39
  \\
  \tracking & 200.00 & 150.30 & 584.44 & 
  205.37 & 243.93 & 276.81
  \\
  \ippo & 0.00 & 25.30 & 137.78 & 38.01 & 12.88 & 42.79 \\
  \mappo & 30.00 & 16.27 & 60.00 & 44.21 & 19.70 & 34.04
  \\
  \bottomrule
  \end{tabular}
\end{table*}

\begin{table*}[t]
  \centering
  \caption{
  The total reward results of the models are presented, following the descriptive style as in Table~\ref{tab:results_med}. Some benchmarks are omitted to avoid repetition.
  } 
  \label{tab:results_high}
  \centering
  \begin{tabular}{lccccc|c}
  \toprule
  \textbf{Model(Reasoning effort:High)} & \textbf{Const-Uni} & \textbf{Dec-Div} & \textbf{Dec-Uni} & \textbf{Inc-Div} & \textbf{Inc-Uni} 
  & \textbf{Average} 
  \\
  \midrule
  \invagent(w/ step desc) & 3.33 & \textbf{48.80} & 206.67 & 147.93 & 265.90 & \textbf{134.53}
  \\
  \invagent(w/ step desc and SS strategy) & \textbf{0.00} & 129.22  & 342.22  & 344.63  & 437.88  & 250.79
  \\
  \ourage(w/o RL log) & 36.67 & 80.48 & 185.33 & 130.90 & 384.55 & 163.59
  \\
  \ourage(w/ RL log) & 19.33 & 50.90 & 188.89 & \textbf{104.13} & 390.45 & 150.74
  \\
  \invagent (w/o strategy) & 46.67 & 66.87 & \textbf{177.78} & 412.81 & 800.00 & 300.83
  \\
  \invagent (w/ strategy) & 190.00 & 68.07 & 666.67 & 127.27 & \textbf{151.51} & 240.70
  \\
  \bottomrule
  \end{tabular}
\end{table*}

\begin{remark}
{\rm 
The complexity of problemdifficulty setting  varies across experimental scenarios, which can explain why the models succeed in some scenarios but not in others. One of our motivations for this scenario configuration is to investigate whether \ourage can adapt to these differences.  
For example, both the increasing and decreasing setting stages require selling 54 units. Thus, in a uniform setting, the total initial inventory across all stages is insufficient, requiring manufacturers to purchase raw materials to prevent backlog. However, in diverse settings, manufacturers do not face this requirement. 
}
\end{remark}

\subsection{Experimental Results and Discussion}
This section presents the experimental results. First, we discuss the overall outcomes, followed by a detailed analysis of the results for each scenario. Finally, we discuss the effects of the reasoning effort. Additional numerical results, including those with standard deviations and those obtained using GPT-5, are presented in Appendix~\ref{appendix:numerical}.

\subsubsection{Overall Results}
Tables~\ref{tab:results_med} and ~\ref{tab:results_high} present the results of the o4-mini-based MASs with medium and high reasoning efforts, respectively.
These tables present the relative gap in the average total rewards $\Delta = |(\opt - r)/\opt|$, where \opt is the optimal value and $r$ is the average total reward.

As shown in these tables,  
the safety-stock strategy enabled the o4-mini-based MAS to achieve optimal performance in the constant demand scenario. 
However, in the scenarios involving time-varying demand, the safety-stock strategy was insufficient 
and proved to have limited effects. 
This limitation resulted from the safety-stock calculation, which generally relies on historical demand data for its predictions, making the system unable to adapt to changes in emerging demand trends. 
In contrast, \ourage (w/ RL log) with both medium and high reasoning efforts achieved high average rewards among all models. 
Notably, with medium reasoning effort, it ranked first across all experimental scenarios and achieved a performance comparable to those of RL benchmark methods, \ippo\ and \mappo, in terms of the average reward. 
A comparison between \ourage(w/ RL log) and \ourage(w/o RL log) further demonstrates that incorporating RL logs into \ourage improved performance, which indicates that \ourage used the logs appropriately. 
These findings suggest that \ourage (w/ RL log) can offer robust adaptability and effectiveness, making it suitable for various  supply chain situations.

\subsubsection{Constant-Uniform}
\begin{figure*}[!t]
  \includegraphics[width=1.0\textwidth]{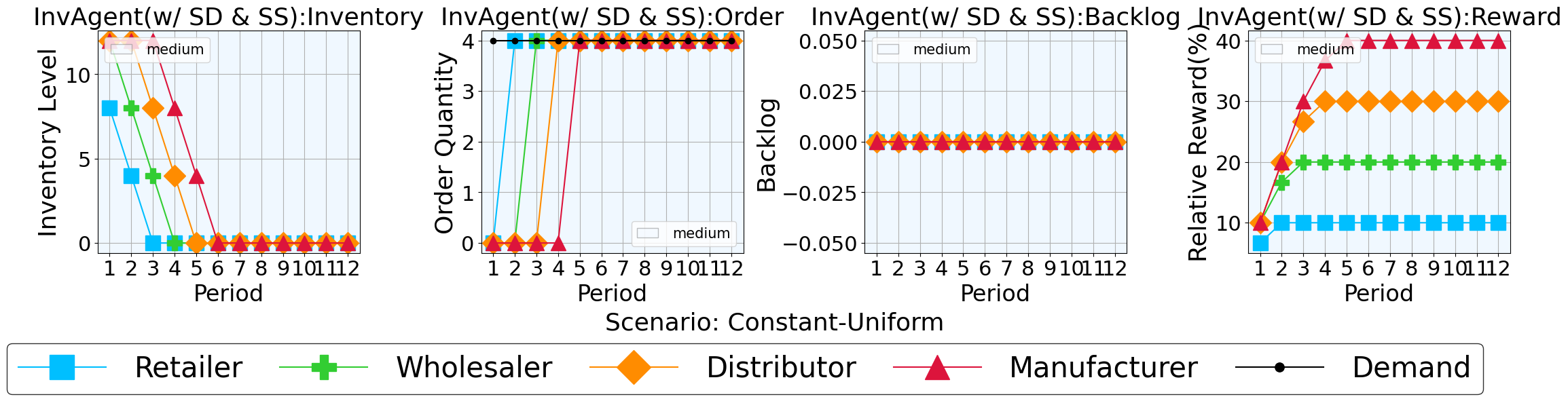}
  \caption{Results of \invagent (w/ step desc and safety stock strategy) with medium reasoning effort. These four graphs present, from left to right, the inventory, backlog, order quantities, and relative reward $(r / \opt)$ within an episode. The “demand” legend corresponds to customer demand, which is plotted in the graph for order quantity.
  }
  \Description{The results of Const-Uni}
  \label{fig:detailedresults}
\end{figure*}

In the constant-uniform scenario, as shown in Tables~\ref{tab:results_med} and~\ref{tab:results_high}, the combination of the step description and safety-stock strategy yields the optimal reward whereas the step description alone also yields strong performance. 
The detailed behavior of \invagent (with step descriptions and strategy) is shown in Figure~\ref{fig:detailedresults}. As shown in the graphs, each agent within the supply chain maintain both inventory and backlog at zero by precisely matching incoming shipments to sales, thereby minimizing the holding costs. Notably, all agents successfully grasped the stage structure and consistently executed an optimal strategy by placing orders for four units at the best time for their respective positions. 
In contrast, other models, including \mappo, did not consistently identify the optimal order quantities. 

\begin{figure*}
  \includegraphics[width=1.0\textwidth]{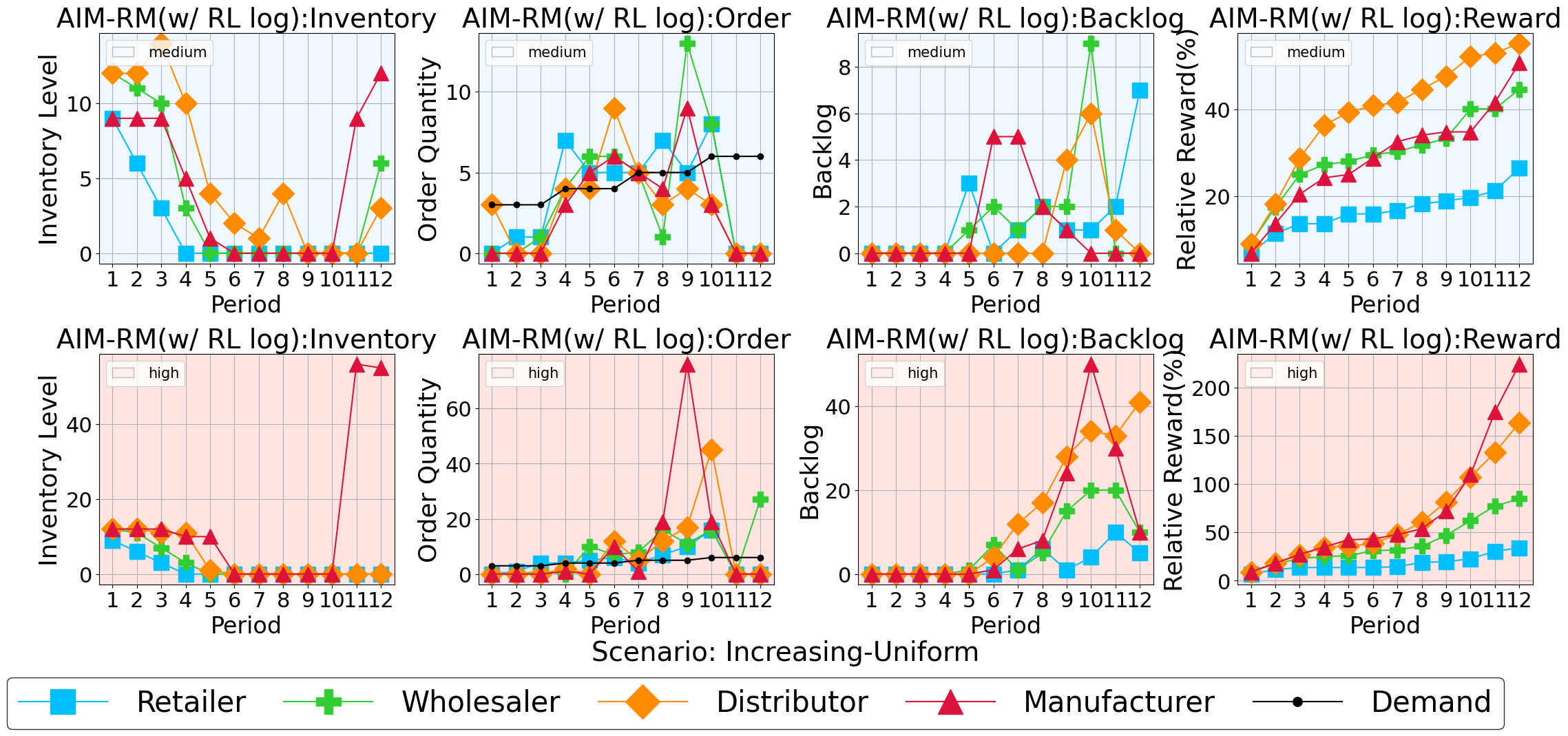}
  \caption{Results for the inventory level, backlog, order quantities, and cumulative reward within an episode.  
  The graphs in the first line represent the results of \ourage (w/ RL log) with medium reasoning effort, and those in the second line represent the same model with a high reasoning effort. 
  }
  \Description{Results of Inc-Uni}
  \label{fig:detailedresults_incuni}
\end{figure*}

\subsubsection{Decreasing-Diverse Scenario} 
As shown in Tables~\ref{tab:results_med} and ~\ref{tab:results_high}, \ourage (w/ RL log) outperforms the other benchmarks, ranking first with medium reasoning effort and second with high reasoning effort. 
The strong performance stems from the agents' ability to review historical logs, enabling them to place orders based on both the logs and  demand trend. 
Furthermore, the agent at each tier successfully places orders by considering the individual lead time, resulting in consistent ordering behavior across all tiers and overall superior performance. 
The tables show that for \invagent(w/o strategy), the performance improves with increased reasoning effort, eventually becoming comparable to that of \ourage(w/ RL log). This is because \invagent(w/o strategy) with high reasoning effort successfully determines order quantities based on specific numbers for future demand forecasts and lead times. In contrast, with medium reasoning effort, the agents do not place orders based on specific numbers.

\subsubsection{Decreasing-Uniform Scenario} 
In the Decreasing-Uniform scenario, as shown in Tables~\ref{tab:results_med} and ~\ref{tab:results_high}, \ourage(w/ RL log) generally outperformed the other benchmarks, ranking first with a medium reasoning effort and third with a high reasoning effort. 
In particular, with high reasoning, \invagent(w/o strategy) slightly outperformed \ourage(w/ RL log).

\subsubsection{Increasing-Diverse Scenario}
In the Increasing-Diverse scenario, as shown in Tables~\ref{tab:results_med} and ~\ref{tab:results_high}, \ourage(w/ RL log) outperforms the other benchmark models with both medium and high reasoning efforts.
The superior performance results from the same mechanism observed in the Decreasing-Diverse scenario: using RL logs enables agents to flexibly place orders based on historical data and the current situation. 
In contrast, \invagent(w/ step description) performed worse as the reasoning effort increases. 
This decline in performance resulted from an underestimation of order quantities downstream of the supply chain, an approach aimed at minimizing the holding costs. The combination of this underestimation and the increasing demand trend results in abrupt order surges as backlog accumulates, triggering the bullwhip effect and ultimately decreasing the reward.

\subsubsection{Increasing-Uniform Scenario}
In the Increasing-Uniform scenario, as shown in Table~\ref{tab:results_med}, \ourage(w/ RL log) achieved the highest performance with a medium reasoning effort. 
However, as indicated in Table~\ref{tab:results_high}, 
its performance declined with higher reasoning effort. 
Figure~\ref{fig:detailedresults_incuni} illustrates that \ourage with high reasoning effort maintained lower inventories, reducing holding costs more than \ourage with medium reasoning effort, thereby causing inventory shortages and sudden large orders at upstream tiers, particularly at the Distributor and Manufacturer. This behavior leads to the bullwhip effect.
In contrast, \invagent (w/ strategy) placed large orders at all tiers without any numerical support, which accidentally matched the increasing demand and caused smaller backlogs. Consequently, the total reward is better than that of the \ourage variants.

\subsubsection{Effects of Reasoning Effort}
Our results demonstrate that increasing reasoning effort does not consistently improve performance. Instead, its effect depends on the complexity of the prompt. 
As shown in Tables~\ref{tab:results_med} and ~\ref{tab:results_high}, 
both \ourage (w/ RL log) and \ourage (w/o RL log), which have complex prompts, showed a decrease in performance as reasoning effort increases across nearly all scenarios. 
In contrast, for relatively simple prompt cases, both \invagent (w/ strategy) and \invagent (w/o strategy) using o4-mini with medium reasoning effort outperform the \invagent variants using GPT-4.1. 
Furthermore, in these simple prompt settings, the \invagent variants generally performed better with high reasoning effort than with medium reasoning effort. 
These results suggest that increasing reasoning effort enhances performance with simple prompts but may be detrimental with complex ones: for example, in models such as \invagent\ (w/ step and safety-stock) and \ourage\ (w/ RL log), greater reasoning effort leads to a decline in performance. 
The patterns may result from the effect of {\it overthinking} \cite{ghosal2025does}, 
which is a critical issue in reasoning models and warrants further investigation. 
Our experiment provided new support for this effect.

\section{Conclusion}
In this study, we investigated LLM-based MASs for multi-echelon inventory management, where the agent at each tier attempts to minimize costs through precise ordering. This study had two primary objectives. First, we investigated the current limitations of LLM-based MASs in the field of inventory management. Our experiments revealed that an LLM-based MAS identifies and implements an optimal ordering policy in the constant-uniform setting using a prompt that describes the process of the inventory management problem and a standard safety stock strategy. However, such an MAS fails to perform effectively in different supply chain scenarios, highlighting a limitation of LLM-based MASs.
The second goal was to enhance adaptability. 
To address this goal, we proposed an LLM-based MAS, \ourage, which leverages similar historical experiences through similarity matching. 
Our numerical experiments demonstrated the robustness and adaptability of \ourage, which outperformed different benchmarks and exhibited performance comparable to that of RL methods.

We identified three primary limitations to our study. 
First, the similarity match employs a fixed similarity measure (the Euclidean norm) and a similarity threshold value $\tau=2$.  
The effect of this limitation on performance requires further investigation. 
Second, \ourage (w/ RL log) occasionally seems to place an order without considering the input similarity cases. 
However, reasoning models tend to output the consequent result without executing  the intermediate reasoning process \cite{chen2025reasoning}. 
Thus, assessing this problem in detail is difficult within our experimental settings.
Regarding the effect of reasoning effort on performance, we observed that increasing the reasoning effort does not necessarily improve performance. Approaching this effect requires 
intensive investigations across reasoning models other than those of OpenAI, including Gemini 2.5 \cite{comanici2025gemini} and Claude Opus \cite{AnthropicOpus}. However, these investigations are beyondout the scope of the present study.

We identified two promising directions for future research. First, \ourage could be enhanced by integrating the self-reflection mechanism developed in \cite{shinn2023reflexion}.  
This mechanism enables LLMs to produce a brief reflection summarizing what was successful, what failed, and the underlying reasons, as well as create actionable rules for future rounds. 
In particular, by adding past self-reflections to historical logs, \ourage can leverage the self-reflection log in similar cases, resulting in improved performance.
Second, we can extend the customer demand settings to include stochastic demand scenarios . 
The use of deterministic demand settings in this study was motivated by the high API  costs associated with running a statistically sufficient number of trials using LLMs. 
To avoid these high costs, we can now use local high-performance reasoning models, such as gpt-oss-120B \cite{agarwal2025gpt}, enabling more extensive experiments even under stochastic conditions.

\section*{Acknowledgements}
The authors appreciate helpful and constructive comments for the anonymous reviewers.

\clearpage
\bibliography{cleaned}
\bibliographystyle{ACM-Reference-Format} 

\clearpage
\appendix
\section{Details of Numerical Experiments  }

\subsection{Additional Numerical Results}
\label{appendix:numerical}
\paragraph{Optimal Value}
The optimal total revenue across all tiers is obtained by maximizing the total revenue under the constraints from \eqref{eq:replenish} to \eqref{eq:inv}. Solving this optimization problem is feasible only if a centralized decision-maker is present, which represents a stronger condition than the decentralized decision-making setting we considered. 
Thus, achieving optimal policies in a decentralized decision-making system is extremely difficult. 
We solved the optimization problem using the CP-SAT solver from OR-Tools \cite{cpsatlp} in Python. The values of \opt for the experimental scenarios are presented in Table~\ref{tab:optresults}.

\paragraph{Standard Deviation}
The average rewards and their standard deviations are presented in Tables~\ref{tab:results_med_retouch} and ~\ref{tab:results_high_retouch}. The standard deviation was $0$ for nearly all experiments. These results suggest that the stochastic behavior of the LLM output does not significantly affect decision-making. 

\subsection{Results for GPT-5}
In the main body, we conducted experiments using o4-mini. Additionally, we conduct the same experiments using GPT-5 with both medium and high reasoning efforts. 
The results, presented in Tables~\ref{tab:results_med_gpt5} and ~\ref{tab:results_high_gpt5}, respectively, 
 demonstrate that \ourage (w/ RL log) with a medium reasoning effort of o4-mini exhibited the optimal average performance among all the results presented in Tables~\ref{tab:results_med}, ~\ref{tab:results_high}, ~\ref{tab:results_med_gpt5}, and ~\ref{tab:results_high_gpt5}. 
The results for GPT-5 with both medium and high reasoning efforts were comparable in average to that for o4-mini with medium reasoning effort. This implies that the degree of overthinking was moderate with GPT-5. \invagent variants exhibited comparable performance to \ourage (w/ RL log). These trends suggest that advancements in LLMs will enable them to independently make sophisticated decisions, eliminating the need for complex prompt tuning and additional tools.

\begin{table}
  \centering
  \caption{
  This table presents the optimal reward in the experimental scenarios
  } 
  \label{tab:optresults}
  \centering
  \begin{tabular}{lccccc}
  \toprule
  \textbf{Model} & \textbf{Const-Uni} & \textbf{Dec- Div} & \textbf{Dec- Uni} & \textbf{Inc-Div} & \textbf{Inc-Uni} 
  \\
  \midrule
  \opt & -120.00 & 332.00 & -45.00 & 242.00 & -132.00
  \\
  \bottomrule
  \end{tabular}
\end{table}

\begin{table*}[t]
  \centering
  \caption{
  This table presents the average total reward for each model across episodes, with the standard deviation indicated in parentheses.
  The description style is the same as that used in Table~\ref{tab:results_med}. The heuristic and RL benchmarks are omitted.
  } 
  \label{tab:results_med_retouch}
  \centering
  \begin{tabular}{lccccc|c}
  \toprule
  \textbf{Model (Reasoning effort: Medium)} & \textbf{Const-Uni} & \textbf{Dec- Div} & \textbf{Dec- Uni} & \textbf{Inc-Div} & \textbf{Inc-Uni} 
  & \textbf{Average} 
  \\
  \midrule
  \invagent(w/ step desc) & 3.33(0.00) & 79.22(0.00) & 437.78(0.00) & 104.55(0.00) & 137.88(0.00) & 152.55
  \\
  \invagent(w/ step desc and SS strategy) & \textbf{0.00(0.00)} & 115.06(0.00) & 315.56(0.00) & 264.88(0.00) & 425.00(0.00) & 224.10
  \\
  \ourage(w/o RL\_log) & 69.33(18.31) & 66.27(2.32) & 219.11(13.38) & 162.64(49.32) & 176.36(57.38) & 138.74
  \\
  \ourage(w/ RL\_log) & 60.00(11.74) & \textbf{56.02(0.00)} & \textbf{171.11(0.00)} & \textbf{95.12(31.03)} & \textbf{74.09(3.90)} & \textbf{91.27}
  \\
    \invagent (GPT-4.1, w/o strategy) & 120.00(0.00) & 244.58(0.00) & 991.11(0.00) & 474.79(0.00) & 262.88(0.00) & 418.67 \\
  \invagent (GPT-4.1, w/ strategy) & 76.67(0.00) & 88.25(0.00) & 428.89(0.00) & 133.06(0.00) & 181.81(0.00) & 181.74 \\
  \invagent (w/o strategy) & 46.67(0.00) & 100.30(0.00) & 202.22(0.00) & 171.90(0.00) & 93.18(0.00) & 122.85
  \\
  \invagent (w/ strategy) & 183.33(0.00) & 118.67(0.00) & 200.00(0.00) & 179.33(0.00) & 155.30(0.00) & 167.33
  \\
  \bottomrule
  \end{tabular}
\end{table*}

\begin{table*}[t]
  \centering
  \caption{
  The results of the total rewards of the models. 
  The number within parentheses represents the standard deviation. The description style is the same as that used in Table~\ref{tab:results_med}. The heuristic and RL benchmarks are omitted.
  } 
  \label{tab:results_high_retouch}
  \centering
  \begin{tabular}{lccccc|c}
  \toprule
  \textbf{Model(Reasoning effort:High)} & \textbf{Const-Uni} & \textbf{Dec- Div} & \textbf{Dec- Uni} & \textbf{Inc-Div} & \textbf{Inc-Uni} 
  & \textbf{Average} 
  \\
  \midrule
  \invagent(w/ step desc) & 3.33(0.00) & \textbf{48.80(0.00)} & 206.67(0.00) & 147.93(0.00) & 265.90(0.00) & 134.53
  \\
  \invagent(w/ step desc and SS strategy) & \textbf{0.00(0.00)} & 129.22 (0.00) & 342.22 (0.00) & 344.63 (0.00) & 437.88 (0.00) & 250.79
  \\
  \ourage(w/o RL log) & 36.67(5.58) & 80.48(14.94) & 185.33(6.22) & 130.90(30.68) & 384.55(18.63) & 163.59
  \\
  \ourage(w/ RL log) & 19.33(1.33) & 50.90(0.00) & 188.89(0.00) & \textbf{104.13(0.00)} & 390.45(8.18) & \textbf{150.74}
  \\
  \invagent (w/o strategy) & 46.67(0.00) & 66.87(0.00) & \textbf{177.78(0.00)} & 412.81(0.00) & 800.00(0.00) & 300.83
  \\
  \invagent (w/ strategy) & 190.00(0.00) & 68.07(0.00) & 666.67(0.00) & 127.27(0.00) & \textbf{151.51(0.00)} & 240.70
  \\
  \bottomrule
  \end{tabular}
\end{table*}

\begin{table*}[t]
  \centering
  \caption{
  This table presents the average total reward for each model with GPT-5 across episodes. 
    The rewards are expressed as the relative gap $\Delta = |\opt - r|/ \opt$, where $\opt$ is the optimal value and $r$ is the total reward. {\it const-uni} in the first line stands for constant-uniform setting. The other names in the first line are also abbreviations of the setting names.
  } 
  \label{tab:results_med_gpt5}
  \centering
  \begin{tabular}{lccccc|c}
  \toprule
  \textbf{Model(Reasoning effort:medium)} & \textbf{Const-Uni} & \textbf{Dec- Div} & \textbf{Dec- Uni} & \textbf{Inc-Div} & \textbf{Inc-Uni} 
  & \textbf{Average} 
  \\
  \midrule
  \invagent(w/ step desc) & 23.33(0.00) & 65.06(0.00) & 202.22(0.00) & 119.01(0.00) & 170.68(0.00) & 116.06
  \\
  \invagent(w/ step desc 
  \\and our strategy) & \textbf{0.00(0.00)} & 60.84 (0.00) & 282.22 (0.00) & 376.45 (0.00) & 356.39 (0.00) & 195.18
  \\
  \ourage(w/o RL log) & 17.33(1.33) & 55.18(3.80) & 175.11(8.00) & 147.11(11.57) & 129.55(5.25) & 104.86
  \\
  \ourage(w/ RL log) & 13.33(0.00) & 51.81(0.00) & 226.67(0.00) & 118.60(0.00) & 112.03(0.00) & 104.49
  \\
  \invagent (w/o strategy) & 46.67(0.00) & \textbf{31.33(0.00)} & \textbf{173.33(0.00)} & \textbf{102.48(0.00)} & 115.04(0.00) & \textbf{93.77}
  \\
  \invagent (w/ strategy) & 133.33(0.00) & 131.33(0.00) & 340.00(0.00) & 108.68(0.00) & \textbf{81.20(0.00)} & 158.91
  \\
  \bottomrule
  \end{tabular}
\end{table*}

\begin{table*}[t]
  \centering
  \caption{
  The results of the total reward with GPT-5 using a high reasoning effort.  
  The description style is the same as that used in Table~\ref{tab:results_med}. The heuristic and RL benchmarks are omitted.
  } 
  \label{tab:results_high_gpt5}
  \centering
  \begin{tabular}{lccccc|c}
  \toprule
  \textbf{Model (Reasoning effort: high)} & \textbf{Const-Uni} & \textbf{Dec- Div} & \textbf{Dec- Uni} & \textbf{Inc-Div} & \textbf{Inc-Uni} 
  & \textbf{Average} 
  \\
  \midrule
  \invagent(w/ step\_desc) & 76.67(0.00) & \textbf{40.06(0.00)} & \textbf{169.33(18.51)} & \textbf{95.04(0.00)} & 191.67(0.00) & 114.55
  \\
  \invagent(w/ step\_desc 
  \\and our\_strategy) & \textbf{0.00(0.00)} & 73.80(0.00) & 202.02(0.00) & 264.46(0.00) & 444.70(0.00) & 197.04
  \\
  \ourage(w/o RL\_log) & 23.33(0.00) & 42.11(1.69) & 175.11(0.89) & 114.38(5.62) & 158.64(19.09) & 102.71
  \\
  \ourage(w/ RL\_log) & 28.67(2.67) & 44.88(0.00) & 195.56(0.00) & 102.07(0.00) & 134.24(2.46) & 101.08
  \\
  \invagent (w/o strategy) & 73.33(0.00) & 51.20(0.00) & 173.33(0.00) & 111.98(0.00) & 71.97(0.00) & \textbf{96.36}
  \\
  \invagent (w/ strategy) & 103.33(0.00) & 93.37(0.00) & 388.89(0.00) & 170.25(0.00) & \textbf{64.39(0.00)} & 164.05
  \\
  \bottomrule
  \end{tabular}
\end{table*}

\subsection{HyperParameter Tuning of RL Comparison Targets}
\label{appendix:reinforce}
We determined the hyperparameters using the same method as described in \cite{quan2024invagent}. We explored the hyperparameters using the candidate values listed in Table~\ref{tab:hypercandidates}. 
We randomly selected 30 combinations of hyperparameters from the table, including the hidden layer size, learning rate, and batch size, and we selected the combination that achieved the highest score. 
The selected hyperparameters of \ippo~and \mappo~ for each experimental scenario are listed in Tables~\ref{tab:IPPOhyper} and ~\ref{tab:MAPPOhyper}, respectively.
Using these selected hyperparameters for each experimental scenario, we evaluated the rewards, as shown in Tables~\ref{tab:results_med} and ~\ref{tab:results_high}.

\begin{table*}[!t]
  \centering
    \caption{ Candidates values of hyperparameters in RL benchmarks.}
    \label{tab:hypercandidates}
  \begin{tabular}{lc}
    \toprule
    Hyperparameter & Candidate Values
    \\
    \midrule
    Number of Hidden Layers & [[128, 128], [256, 256]]
    \\
    Activation Function& [ReLU] 
    \\
    Learning rate& [1e-4, 5e-4, 1e-3] 
    \\
    Learning Batch Size& [500, 1000, 2000] 
    \\
    SGD Minibatch Size& [32, 64, 128] 
    \\
    Number of SGD Iterations& [5, 10, 20] 
    \\
    Number of  Learning Iterations& [500, 800, 1000, 1500] 
    \\
    $\gamma$ & [1.0] 
    \\
    \bottomrule
  \end{tabular}
\end{table*}

\begin{table*}[!t]
  \centering
\caption{Hyperparameters of \ippo~in each scenario.}
  \label{tab:IPPOhyper}
  \begin{tabular}{lccccc}
    \toprule
    {\bf Hyperparameter} & {\bf Const-Uni} & {\bf Dec-Div} & {\bf Dec-Uni} & {\bf Inc-Div} & {\bf Inc-Uni }
    \\
    \midrule
    Numbers of Hidden Layer & [128, 128] & [128, 128] & [256, 256] & [128, 128] & [256, 256]
    \\
    Activation Function& ReLU & ReLU & ReLU & ReLU & ReLU
    \\
    Learning rate& 1e-4 &1e-4,& 1e-4 & 1e-4 & 1e-4 
    \\
    Training Batch Size&  1000 &1000 & 1000 & 500 & 1000
    \\
    SGD Minibatch Size & 128 &128& 128 &32 & 128
    \\
    Numbers of SGD Trials& 5 & 5& 5& 10& 5
    \\
    Numbers of  Learning Trials& 1000 & 1000& 1500 & 800 & 1500 
    \\
    $\gamma$ & 1.0 & 1.0 & 1.0 & 1.0 & 1.0 
    \\
    \bottomrule
  \end{tabular}
\end{table*}

\begin{table*}[!t]
  \centering
    \caption{Hyperparameters of \mappo~in each scenario.}
    \label{tab:MAPPOhyper}
  \begin{tabular}{lccccc}
    \toprule
    {\bf Hyperparameter} & {\bf Const-Uni}& {\bf Dec-Div} & {\bf Dec-Uni} & {\bf Inc-Div} & {\bf Inc-Uni}
    \\
    \midrule
    Numbers of Hidden Layer & [256, 256] & [128, 128] & [128, 128] & [128, 128] & [256, 256]
    \\
    Activation Function& ReLU & ReLU & ReLU & ReLU & ReLU
    \\
    Learning rate& 1e-4 &1e-4& 1e-4 & 1e-4 & 1e-4 
    \\
    Training Batch Size& 1000 &1000 & 1000 & 500 & 1000
    \\
    SGD Minibatch Size & 128 &64& 32 &128 & 64
    \\
    Numbers of SGD Trials& 5 & 5& 5& 10& 5
    \\
    Numbers of  Learning Trials& 1500 & 1000& 1000 & 500 & 1000 
    \\
    $\gamma$ & 1.0 & 1.0 & 1.0 & 1.0 & 1.0 
    \\
    \bottomrule
  \end{tabular}
\end{table*}

\section{Prompt Examples}
\label{appendix:prompts}

In our numerical experiments, we used several types of prompts, each illustrated in the figures referenced below. Figure~\ref{pro:decision} presents an example of a decision prompt $\decisionprompt$, whereas Figure~\ref{pro:step} displays a step-wise description prompt \stepp. The safety stock description \safestockp is provided in Figure~\ref{pro:saftystock}, and the memory usage prompt \memoryp is illustrated in Figure~\ref{pro:memory}. Finally, the demand description employed in the experiments is shown in Figure~\ref{pro:demand}. Furthermore, in this section, we added the details of the prompts.

\paragraph{Decision Prompt}
This prompt guides an agent by indicating its location and current state, including the lead time, inventory level, and current backlog. Using these inputs, the agent essentially decides an action through this prompt. 
However, this prompt does not include any information about the problem setting or strategies for inventory management.

\paragraph{Step-Wise Prompt}
This prompt outlines the procedure executed during a period, as described in Section~\ref{sec:agentbehavior}, including an example of lead-time calculation.   
This procedure is crucial for calculating the required inventory level and subsequently enabling an agent to precisely select an action along with the prompt. It is particularly crucial when the order issued by the agent arrives or the inventory is consumed.

\paragraph{Safety-Stock Prompt}
This prompt teaches agents a standard strategy for inventory management. The strategy involves first calculating the inventory position by considering the lead time, which refers to the future inventory after the lead time. Subsequently, the strategy calculates the target inventory level, which is the quantity expected to be consumed  by the end of the lead time. The quantity to be ordered can be calculated based on the gap between the previous two quantities. In particular, the strategy explicitly indicates the number of periods required to calculate the total arrival demand, which enables an agent to appropriately decide an order quantity in the Const-Uni setting.

\paragraph{Memory Usage Prompt}
This prompt defines the term 'similar case,' which comprises a state vector, action, reward, and distance (similarity measure).
It also teaches the agent the meaning and positioning of similar cases. 
These similar cases are input to the agent by substituting them into this prompt.

\paragraph{Demand Description}
The demand description outlines the customer demand that appears in each period. This study  assumes a deterministic demand, meaning that the described demand will certainly appear.

\clearpage
\begin{figure*}[t]
\centering
    \begin{tcolorbox}
    {\bf Decision prompt}:
    \\
Now this is the round {period}, and you are at the stage {stage} of \{num\_stages\} in the supply chain. Given your current state:
    \\
    - Lead Time: {lead\_time} round(s)
    \\
    - Inventory Level: {inventory} unit(s)
    \\
    - Current Backlog (you owing to the downstream): {backlog} unit(s)
    \\
    - Upstream Backlog (your upstream owing to you): {upstream\_backlog} unit(s)
    \\
    - Previous Sales (in the recent round(s), from old to new): {sales}
    \\
    - Arriving Deliveries (in this and the next round(s), from near to far): \{deliveries\}"
    \\
\{demand\_description\}
\\
Your downstream order from the stage \{stage\_idx\} for this round is \{Action(stage\_id-1)\}.
    \end{tcolorbox}
    \caption{This prompt teaches the decision-making module input data including the current state and the output style.}
    \Description{Decision Prompt}
    \label{pro:decision}
\end{figure*}

\begin{figure*}[!t]
\centering
    \begin{tcolorbox}
{\bf Step Description}: \\
\#\#\#\#\#\#\#\#\#\#\#\#\#\#\#\#\#\#\#\#\#\#\#\#\#\#\#\#\#\#\#\#\#\#\#\#\#\#\#\#\#\#\#\#\#\#\#\#\#\#\#\#
\\
\#\#\#\#\#\#\#\#\#\#\#\#\#\#\#\#\#\
\\
\#\#\#   One-Period Flow in InventoryManagementEnv (as implemented)  
\#\#\#
\\
\#\#\#\#\#\#\#\#\#\#\#\#\#\#\#\#\#\#\#\#\#\#\#\#\#\#\#\#\#\#\#\#\#\#\#\#\#\#\#\#\#\#
\#\#\#\#\#\#\#\#\#\#\#\#\#\#\#\#\#\#\#\#\#\#\#\#\#\#\#

1. A period has the following four steps:
\\
a) Receive delivery: Receive deliveries ordered previously and update inventory levels.
\\
b) Order decision: Decide on the order quantity to be placed with upstream suppliers.
\\
c) Ship items: Ship items ordered this period to downstream suppliers.
\\
d) Calculate profits: Calculate profits based on sales, purchases, and costs.
\\
Lead-time example (L\_m = 2)
  self.period  :  1   2   3   4  …\\
  see inventory:  I0  I1  I2* I3 …\\
  decide order :  O1  O2  O3  …\\
  \ \ \ \ \                 {$\hookdownright$} ship R1\\
                               {$\hookdownright$} arrives → added to I3 
                               \\
  *During the decision at t = 3, the lot ordered at t = 1 has not yet arrived.
    \end{tcolorbox}
    \caption{The \stepp teaches a LLM the procedure of the multi-echelon inventory management.}
    \label{pro:step}
    \Description{Step prompt}
\end{figure*}

\begin{figure*}
    \begin{tcolorbox}
    {\bf Safety-stock Description}: 
    \\
    **Decision Policy (LT-safe order-up-to) — follow strictly**
    \\
1) Compute your Inventory Position (IP) taking lead time into account:
IP = inventory + sum(deliveries over the next lead\_time periods) - current\_backlog.
\\
2) Adopt a periodic-review order-then-ship order-up-to policy.
Set the base target as: target\_base $\approx$ (lead\_time + 1) × $\mu\hat{}$(mu\_hat),
and optionally add safety stock: z × $\sigma\hat{}$ (sigma\_hat) × $\sqrt{}$(lead\_time + 1).
\\
3) Determine the order quantity by considering the gap between target\_base and your current IP. However, ensure that the order quantity does not exceed the production capacity ({prod\_capacity}).
\\
4) Provide a brief rationale (1–2 sentences) explaining how you arrived at the order quantity.
    \end{tcolorbox}
    \caption{This prompt describes a standard safety stock strategy in Manufacturing.}
    \label{pro:saftystock} 
    \Description{Safety-stock prompt}
\end{figure*}

\begin{figure*}[!t]
    \centering
    \begin{tcolorbox}
    {\bf Memory Usage Description}: 
    \\
    You are also given retrieval results from prior episodes, called `similar\_cases`.
    \\
    Definition - `similar\_cases` are historical decisions from prior episodes whose numeric states are most similar to the current state.
    \\
    - State vector fields (in this exact order): [inventory, backlog, upstream\_backlog, lead\_time, deliveries]
    \\
    - Each case includes: `state\_vec`, `action`, `reward`, and `distance` (similarity measure).
    \\
    - Lower distance values indicate higher similarity to your current situation.
    \\
    - Treat them as evidence, not rules.
    similar\_cases:
    \{similar\_cases\}
    \end{tcolorbox}
    \caption{This prompt instructs the decision-making module how to use the similarity cases.}
\label{pro:memory}
\Description{Memory Usage prompt}
\end{figure*}

\begin{figure*}[t]
    \begin{tcolorbox}
    {\bf Demand description}: 
    \\
    {\bf Increasing Demand}: The expected demand at the retailer (stage $0$) is $(2 + \lceil t / 3\rceil )$ for each round t = 1,…,12. Hence, demand progresses in three-round steps, taking the values 3 units in rounds 1–3, 4 units in rounds 4–6, 5 units in rounds 7–9, and 6 units in rounds 10–12.
    \\
    {\bf Decreasing Demand}: The expected demand at the retailer (stage $0$) is $(2 + \lceil(12 - (t - 1)) / 3 \rceil$ ) for each round t = 1,…,12. Accordingly, demand declines in three-round steps, taking the values 6 units in rounds 1–3, 5 units in rounds 4–6, 4 units in rounds 7–9, and 3 units in rounds 10–12.
    \end{tcolorbox}
    \caption{This prompt describes indicates that the customer demand appears at all rounds.}
    \Description{Demand Description}
    \label{pro:demand}
\end{figure*}

\end{document}